\documentclass[graybox,envcountchap]{svmult}

\usepackage{mathptmx}        
\usepackage{amsmath}
\usepackage{amssymb}
\usepackage{color}
\usepackage{helvet}          
\usepackage{courier}         
\usepackage{dirtree}

\usepackage{makeidx}        
\usepackage{graphicx}        
\usepackage{subfig}

\usepackage{multicol}        
\usepackage[bottom]{footmisc}

\usepackage{hyperref}        
\hypersetup{colorlinks=true,urlcolor=blue}

\usepackage[misc]{ifsym}
\usepackage{braket}

\newcommand{\be}{\begin{eqnarray}}
\newcommand{\ee}{\end{eqnarray}}
\newcommand{\pro}[2]{\mbox{$\langle\, #1 \mid #2\,\rangle$}}
\newcommand{\expec}[1]{\mbox{$\langle\, #1\,\rangle$}}

\renewcommand{\a}{\hat a}
\newcommand{\ac}{\hat a^{\dagger}}

\renewcommand{\d}{\mbox{${\rm d}$}} 
\newcommand{\lp}{\ell_{\rm p}}
\newcommand{\mpl}{m_{\rm p}}
\newcommand{\gn}{G_{\rm N}}
 
\newcommand{\rh}{r_{\rm H}}

\newcommand{\Rh}{R_{\rm H}}

\newcommand{\Ng}{{N_{\rm G}}}

\newcommand{\Rinf}{{R_{\infty}}}
\newcommand{\dd }{{\mathrm d}}

\makeindex             

\begin{document}


\title*{Quantum matter core of black holes (and quantum hair)}
\author{Roberto Casadio and Octavian Micu}
\institute{Roberto Casadio (\Letter) \at Department of Physics and Astronomy ``A. Righi'',
Bologna University, via Irnerio 46, 40126 Bologna, Italy, 
\email{casadio@bo.infn.it}
\and Octavian Micu \at Institute of Space Science, P.O.~Box MG-23, 077125 Bucharest-Magurele, Romania, 
\email{octavian.micu@spacescience.ro}}
%
%
\maketitle
\abstract{The idea that gravity can act as a regulator of ultraviolet divergences 
is almost a century old and has inspired several approaches to quantum gravity.
In fact, a minimum Planckian length can be shown to emerge from the nonlinear dynamics of gravity
in the effective field theory approach to gravitational scatterings at Planckian energies.
A simple quantum description of the gravitational collapse of a ball of dust supports the conclusion that such a
length scale is indeed dynamical and matter inside black holes forms extended cores of macroscopic size.
The geometry of these quantum black holes can be described by coherent states which cannot contain modes of arbitrarily
short wavelength, compatibly with a matter core of finite size.
Therefore, the classical central singularity is not realised, yet the expected general relativistic behaviour can be recovered
in the weak-field region outside the horizon with good approximation.
Deviations from classical general relativistic solutions are still present and form quantum hair which
modify the thermodynamical description of black holes.
These quantum black holes also avoid the presence of inner (Cauchy) horizons, since the effective energy density and
pressures are integrable, as required by quantum physics, and not as regular as in classical physics.}
%
%
%
\vspace{1cm}
\noindent
{Invited chapter for the edited book {\it New Frontiers in Gravitational Collapse and Spacetime Singularities}
(Eds. P. Joshi and D. Malafarina, Springer Singapore, expected in 2023)}
\section{Introduction}
\label{sec:1}
Quantum gravity is generically expected to eliminate the singularities predicted by general relativity~\cite{HE}
and several approaches to quantum gravity~\cite{kiefer} have been proposed in this framework.
It is suggestive to recall that the Planck scale corresponds to a black hole of mass $M$ whose horizon radius $\Rh=2\,\gn\,M$ is
of the order of its Compton length $\lambda_M=\hslash/M$.~\footnote{We shall always use units with $c=1$ and often
write the Planck constant $\hslash=\lp\,\mpl$ and the Newton constant $\gn=\lp/\mpl$, where $\lp$ and $\mpl$ are the
Planck length and mass, respectively.}
This simple observation, for example, lies at the heart of generalised uncertainty principles~\cite{bronstein}
which modify the uncertainty relations of quantum mechanics in order to include gravity in scattering processes 
associated with measurements (like in the famous Heisenberg microscope~\cite{heisenberg}).
The idea that gravity involves a minimum (fundamental) length $\lp$ was also viewed as a possible cure for the
ultraviolet divergences of quantum field theory~\cite{pauli}, and regained notoriety with the increasing interest
in quantum gravity and trans-Planckian effects~\cite{Hossenfelder:2012jw}.
Indeed, many candidates for quantum gravity exhibit a minimum length, from string theory to loop quantum gravity.
\par
It is consistent with the above picture that a Planck scale minimum length arises from the Feynman
path integral that generates in-out amplitudes including gravity~\cite{Padmanabhan:1985jq},
hence in $S$-matrix elements via the Lehmann-Symanzik-Zimmermann formula.
However, these amplitudes are acausal and complex since they are subjected to Feynman boundary conditions.
A proper minimum spacetime length should instead be real to arbitrary loop orders as pertains to an expectation value.
We will first clarify this point in the effective field theory on a fixed background geometry, with a general argument
that support the existence only of a Planckian screening length in elementary scattering processes as a consequence
of the gravitational dynamics described by the Einstein-Hilbert action~\cite{Casadio:2020hzs,Casadio:2022opg}.
\par
The next question is then whether such a dynamical scale remains of the Planck size in more complex systems, 
like very compact astrophysical objects eventually collapsing to form black holes.
The nonlinearity of general relativity in fact does not guarantee that the above result carries on unaltered 
when one deals with a huge number of massive particles in a bound state.
A hint of the role played by nonlinearity can be obtained from the bootstrapped Newtonian
gravity~\cite{Casadio:2018qeh,Casadio:2019cux}, which is defined by adding self-interaction terms to the Newtonian
Lagrangian density for spherically symmetric compact objects.
The main consequence of this approach can be characterised by modified relations between the mass $M$ and
radius $R$ of stable configurations, and indeed the lack of a Buchdahl limit~\cite{Buchdahl:1959zz} (which requires $R>9\,\gn\,M/4$).
A fully quantum version of it can also be considered~\cite{Casadio:2020mch} and first evidence for a maximum
compactness for black holes is obtained~\cite{Casadio:2020ueb}.
\par
We will next move on to full general relativity and study the Oppenheimer-Snyder model~\cite{OS} of a collapsing
ball of dust.
Since the canonical quantisation of a large number of (dust) particles as (fundamental) field excitations
is beyond our capability, we will quantise this model in analogy with the quantum mechanical treatment of the
hydrogen atom.
In particular, we shall retain the ball radius $R$ as the only observable and show that it admits a (discrete)
spectrum of bound states~\cite{Casadio:2021cbv}.
Furthermore, as a consequence of the nonlinearity of the governing (Einstein) equations, the ground state
will be shown to have a surface area proportional to $M^2$, so that the mass is quantised according to
Bekenstein's area law.
This implies that no singularity ever appears but collapsing matter forms a quantum core of macroscopic
size. 
Of course, the huge simplification in the model will reflect in a similarly large (configuration) entropy for
such cores~\cite{Casadio:2022pla}.
\par
Having established that the collapse does not end in a singularity due to purely quantum effects,
the geometry generated by the core in the ground state can be described by means of a coherent quantum
state~\cite{Casadio:2021eio}.
Regularity of the geometric quantum state is guaranteed by the finite size of the core, which removes
all ultraviolet divergences and causes deviations from the Schwarzschild geometry.
Since these deviations are regulated by the size of the core, one can clearly conceive that their presence
modifies the thermodynamics and Hawking radiation depending on the matter state actually present
inside the horizon~\cite{Calmet:2021stu}.
Moreover, we will argue how a similar description can avoid the emergence of inner Cauchy horizons,
both in spherical symmetry and for rotating systems~\cite{Casadio:2023iqt},
whose presence casts serious doubts about the physical relevance of classical models of regular black holes.
\section{Minimum length scale in the effective field theory of gravity}
\label{sec:minlength}
In the effective field theory of gravity, the metric tensor defining the line element $\d s^2=g_{\mu\nu}\,\dd x^\mu\,\dd x^\nu$
is promoted to an operator, but one assumes the existence of a classical configuration for matter sources and the
corresponding spacetime metric $\bar g_{\mu\nu}$.
Only ``small'' perturbations, below a certain (energy) threshold, around the chosen classical solution are allowed 
and described as excitations of quantum fields.
We can therefore write
\be
g_{\mu\nu}
=
\bar g_{\mu\rho} \left(e^{\sqrt{\frac{32\,\pi\,\lp}{\mpl}}\,h}\right)^\rho_{\ \nu}
\simeq
\eta_{\mu\nu}
+
\sqrt{\frac{32\,\pi\,\lp}{\mpl}}\,h_{\mu\nu}
+
\frac{16\,\pi\,\lp}{\mpl}\,h_{\mu\rho}\,h^{\rho}_{\ \nu}
\ ,
\label{eq:param}
\ee
where we assumed that the background metric $\bar g_{\mu\nu}=\eta_{\mu\nu}$ for simplicity
(results can be generalised to any curved background by adopting normal coordinates).
The Lagrangian governing the dynamics will include the usual Einstein-Hilbert term and
possible quantum corrections, giving rise to a number of propagating modes $i=1,\ldots$, in the
linearised version for small perturbations.
\par
The geometrical distance between two nearby points, say at coordinates $x^\mu$ and
$y^\mu = x^\mu + \dd x^\mu$,  is computed from the expectation value of the above operator
on a suitable quantum state.
Hence, it can be technically obtained from the real in-in correlation function
\be
\ell_\text{in-in}^2(x,y)
=
\braket{0_\text{in} | \dd s^2 | 0_\text{in}}
\ ,
\label{Linin}
\ee
where the ``vacuum'' $\ket{0_{\rm in}}$ is such that the expected classical background expression
$\ell^2 = \eta_{\mu\nu}\,\dd x^\mu\, \dd x^\nu$ is recovered at the leading order $(\lp/\mpl)^0$.
Note that $\ell^2$ goes to zero when $\dd x^\mu$ vanishes at coincident points, whereas terms of order
$(\lp/\mpl)^{1/2}$ are linear in $h_{\mu\nu}$ and vanish everywhere by the usual Fock space construction.
\par
One can now argue in a model-independent way for the absence of a minimum geometrical
distance along physically allowed trajectories~\cite{Casadio:2020hzs,Casadio:2022opg}.
For almost coincident points, Eq.~\eqref{Linin} reads
\be
\ell_{\text{in-in}}^2
&\simeq&
\frac{16\,\pi\,\lp}{\mpl}\,
\bra{0_{\rm in}}h_{\mu\rho}(x)\,h^{\rho}_{\ \nu}(y)\ket{0_\text{in}}\,
\dd x^\mu\, \dd x^\nu
\nonumber
\\
&\simeq&
\frac{16\,\pi\,\lp}{\mpl}\,
G^\text{ret}_{\mu\nu\rho\sigma}(x,y)\,
\dd x^\mu\, \dd x^\nu
\ ,
\label{eq:ds}
\ee
where we assumed that $x^\mu$ is in the future of $y^\mu$.
The retarded propagator is in general given by
\be	
G^\text{ret}_{\mu\nu\rho\sigma}
=
\sum_i \left[-\frac{\theta(x^0-y^0)}{2\,\pi}\,\delta(\ell^2)
+ \theta(x^0-y^0)\,\theta(\ell^2)\,
\frac{m_i\, J_1(m_i\,\ell)}{4\,\pi\,\ell}\right]
\hslash\,P^i_{\mu\nu\rho\sigma}
\ ,
\label{Gf}
\ee
where $m_i$ is the (possibly vanishing) mass of the $i^{\rm th}$ propagating mode and
\be
P^i_{\mu\nu\rho\sigma}
=
\alpha_i\, \eta_{\mu\rho}\,\eta_{\nu\sigma}
+ \beta_i \,\eta_{\mu\sigma}\,\eta_{\nu\rho}
+ \gamma_i\, \eta_{\mu\nu}\,\eta_{\rho\sigma}
\ee
is the most general fourth rank tensor symmetric in $\{\mu\nu\}$ and $\{\rho\sigma\}$, with
parameters $\alpha_i$, $\beta_i$ and $\gamma_i$ depending on the particular gravitational Lagrangian
for the $i^{\rm th}$ mode.
The contraction $P_{\mu\rho\ \ \nu}^{i\ \ \rho}\,\dd x^\mu\, \dd x^\nu\sim \ell^2$
and Eq.~\eqref{eq:ds} in the coincident limit finally yields
\be
\lim_{x\to y}\,
\ell_{\text{in-in}}^2
=
0
\ .
\ee
This result relies solely on the analytic structure of the retarded propagator in position space.
We can therefore conclude that the effective field theory (of gravity) shows no sign of a minimum geometrical length
along the path of test particles, irrespectively of the specific (gravitational) Lagrangian.
\par
By stressing that one of the main roles of the metric is to determine the geodesic motion of test particles,
one can also conclude that free propagation of (quantum) excitations on the background state $\ket{0_{\rm in}}$
is described by the geodesic equation in the (smooth) background $\bar g_{\mu\nu}=\eta_{\mu\nu}$.
At the very least, this is a proof of consistency of the effective field theory which requires that the quantum state $\ket{0_{\rm in}}$
entails the existence of a classical background spacetime.
\subsection{Length scale in scatterings}
\label{S:minlength}
We just used the fact that in-in amplitudes involve retarded propagators.
For cross sections and decay rates, one instead assumes Feynman boundary conditions
corresponding to the scattering of (particle) excitations of the initial state $\ket{0_\text{in}}$
into those of a final state $\ket{0_\text{out}}$.
The Feynman amplitudes for the metric between two nearby points,
\be
\ell_\text{in-out}^2(x,y)
=
\braket{0_\text{out} | \dd s^2 | 0_\text{in}}
\ ,
\ee
can then be used to determine a length scale for interaction processes involving gravity.
In fact, for almost coincident points, one finds 
\be
\ell_{\text{in-out}}^2
&\simeq&
\frac{16\,\pi\,\lp}{\mpl}\,
\bra{0_{\rm out}}h_{\mu\rho}(x)\,h^{\rho}_{\ \nu}(y)\ket{0_\text{in}}\,
\dd x^\mu\, \dd x^\nu
\nonumber
\\
&\simeq&
\frac{16\,\pi\,\lp}{\mpl}\,
G^\text{F}_{\mu\nu\rho\sigma}(x,y)\,
\dd x^\mu\, \dd x^\nu
\ ,
\label{eq:dsF}
\ee
where the Feynman propagator has the general short distance behaviour
\be
G^\text{F}_{\mu\nu\rho\sigma}
\simeq
\sum_i \frac{i\,\hslash\,P^i_{\mu\nu\rho\sigma}}{4\,\pi^2\left(x-y\right)^2}
\ .
\label{Gret}
\ee
Eq.~\eqref{eq:dsF} then yields $\lim\limits_{x\to y}\,\ell_{\text{in-out}}^2\sim
\lim\limits_{\ell\to 0}\,e^{-\frac{\lp^2}{\ell^2}\,|\Sigma|}\,\ell^2=0$
if
$\Sigma\equiv \sum_i(\alpha_i+4\beta_i+\gamma_i) \leq 0$
and
\be
\lim_{x\to y}\,
\ell_{\text{in-out}}^2
\simeq
i\,\frac{2}{\pi}\,\Sigma\,\lp^2\,
\sim
\lp^2
\qquad
{\rm if}
\quad
\Sigma> 0
\ .
\label{eq:zerolength}
\ee
For positive $\Sigma$, one therefore expects the emergence of a finite length scale proportional to
$\lp$ in the modulus of transition amplitudes that will dynamically screen effects in the ultraviolet.
\par
The above result again follows from general properties of the Feynman propagator for metric perturbations
and the information about the details of the gravitational dynamics is contained solely in the parameters $\alpha_i$, $\beta_i$
and $\gamma_i$.
In particular, the massless spin-2 field (the graviton) is the only degree of freedom in general relativity and $\Sigma=4$.
The corresponding length scale is therefore given by
\be
\ell_{\text{in-out}}(x,x)
\simeq
\sqrt{\frac{8}{\pi}} \, \lp
\ .
\label{Lscale}
\ee
Another interesting example is Stelle's theory~\cite{Stelle:1976gc},
whose spectrum contains additional degrees of freedom needed to prove the renormalisability.
In this case, there is a surprising accidental cancelation of the parameters so that $\Sigma=0$ and 
a minimum length scale does not appear in scattering processes.~\footnote{For a discussion of cases with $\Sigma<0$,
see Ref.~\cite{Casadio:2022opg}.}
\par
Renormalisable theories possess no natural scale, whereas non-renormalisable theories contain
an intrinsic scale which is then used as a cut-off to define the effective field theory and its range of validity.
For lengths below this scale, the effective field theory breaks down and ultraviolet phenomena cannot be described without a ultraviolet completion.
General relativity is not renormalisable and we indeed find a minimum scale of the order of $\lp$,
which is precisely the scale used to perform the effective field theory expansion.
On the other hand, Stelle's theory is renormalisable and does not need any intrinsic scale.
This suggests an interesting correspondence between the renormalisability
of a theory (of gravity) and the non-existence of a minimum length scale.
However, in a perspective like the one of the asymptotic safety scenario~\cite{AS}
and classicalization~\cite{Dvali:2010ns,classicalisation}, one might instead argue that the dynamical emergence of
the length scale $\lp$ does not imply the need of a ultraviolet completion but that the (effective) theory is self-complete
and simply rearranges its degrees of freedom so that no new physics appears below $\lp$.
This realises the old idea that $\lp$ acts as a natural (albeit dynamical rather than geometrical)
regulator that removes all ultraviolet divergences~\cite{pauli,Padmanabhan:1985jdl}
and allows to treat general relativity as a fundamental theory of gravity.
From this viewpoint, it is not general relativity that fails at the Planck scale, but rather physics
beyond $\lp$ becomes (operationally) meaningless.
\subsection{Minimum length scale and generalised uncertainty principle}
\label{S:gup}
From the operational point of view, scattering processes are generically involved in quantum mechanical
measurements, as Heisenberg argued in his famous description of the quantum microscope~\cite{heisenberg}.
The position $x_{\rm e}$ and momentum $p_{\rm e}$ of a particle, say an electron, can be determined
by scattering a photon off the electron and measuring properties of the photon after the scattering.
If $\lambda$ is the photon wavelength, the uncertainty in the position of the electron is (at least)
$\Delta x_{\rm e} \simeq \lambda$.
Moreover, the photon carries a momentum $p=h/\lambda$, which is partially transferred to the electron
in an unknown magnitude and direction.
This implies that, just after the scattering, the uncertainty in the electron momentum amounts to
(at most) $\Delta p_{\rm e} \simeq p = h/\lambda$. 
Heisenberg thus concluded that
\be
\Delta x_{\rm e}\, \Delta p_{\rm e}
\simeq
\lambda \cdot \frac{h}{\lambda}
\simeq
h
\ .
\ee         
Later on, Schr\"odiger and Robinson formulated the uncertainty principle for canonically
conjugated variables, like the $x$ and $p$ of a particle, as
\be
\Delta x \,\Delta p \geq \frac{\hslash}{2}
\ ,
\label{hep}
\ee
which is the most common form used at present.
\par 
Heisenberg's heuristic approach paved the way to the formulation of various
generalised uncertainty principles~\cite{Hossenfelder:2012jw,Casadio:2020rsj}
from taking into account the gravitational effects in the photon-particle interaction.
Indeed, if the centre-of-mass energy is sufficiently large to result in the formation of a black hole,
the corresponding horizon radius acts a screening length below which position measurements become
meaningless~\cite{Scardigli:1999jh}.
Generalised uncertainty principles can be mathematically derived from modified quantum mechanical commutators,
and one is tempted to extend such modifications to quantum field theory commutators.
However, we have seen that a minimum length scale can be naturally obtained without modifying the quantum
field theory dynamics, hence without altering the field commutators.
It would look more physically sound that generalised uncertainty principles therefore emerge effectively
in quantum mechanics as the non-relativistic sector of quantum field theory of gravity without modifying
the field propagators {\em ad hoc}.
\begin{figure}[t]
\centering
\includegraphics[width=8cm]{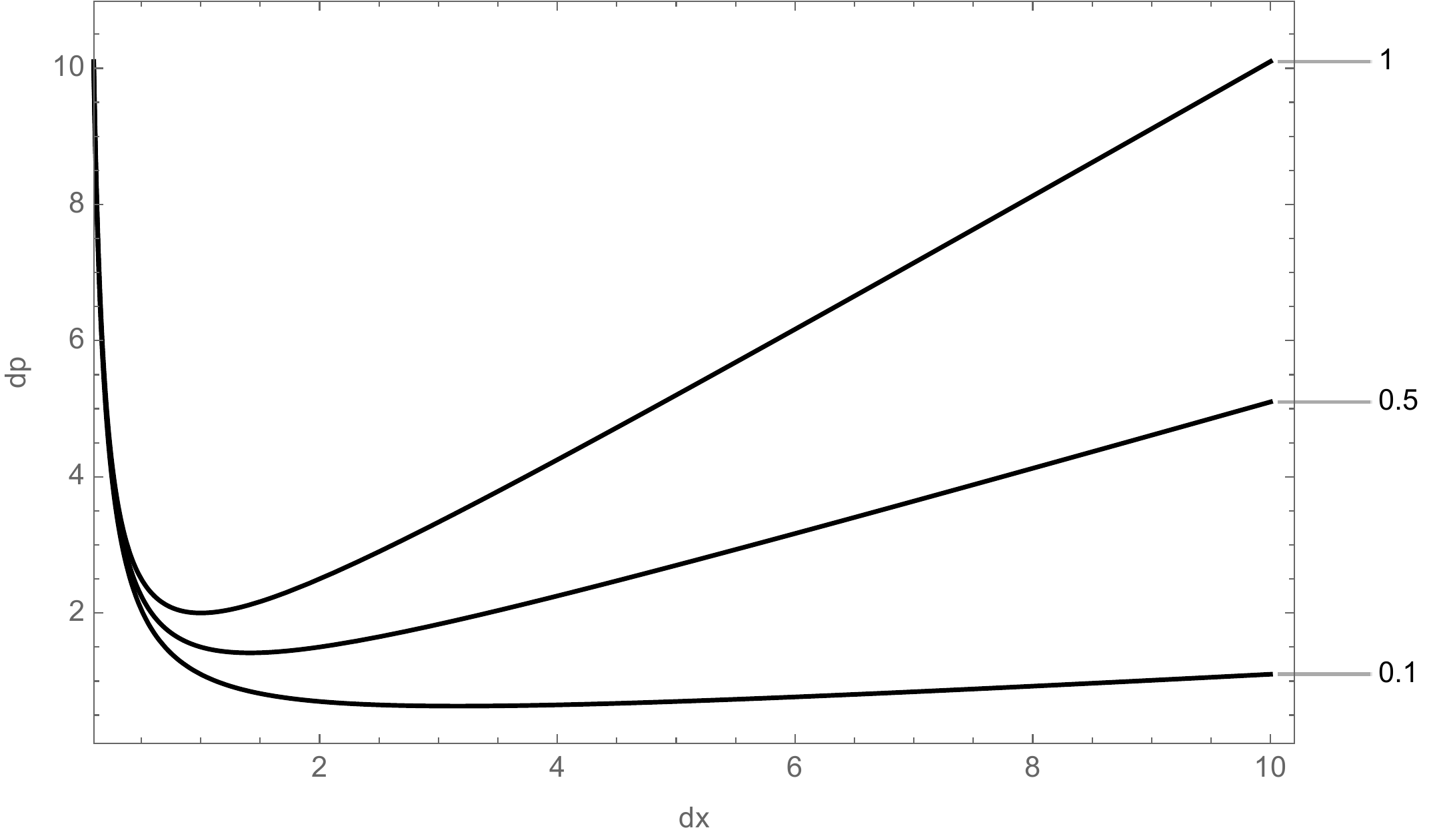}
\caption{Generalised uncertainty principle~\eqref{dxdp} and minimum length scale for different values of $\delta_0$ (lengths and momenta 
in Planck units).}
\label{f:gup}
\end{figure}
\par
The simplest form of generalised uncertainty principle is given by (see Fig.~\ref{f:gup})
\be
\Delta x\,\Delta p
\ge
\frac{\hslash}{2}
\left(1+\frac{\delta_0}{\mpl^2}\,\Delta p^2\right)
\ ,
\label{dxdp}
\ee
where $\Delta x^2\equiv\expec{\hat x^2}-\expec{\hat x}^2$, and similarly $\Delta p^2\equiv\expec{\hat p^2}-\expec{\hat p}^2$,
and $\delta_0$ is a dimensionless deforming parameter expected to emerge
from candidate theories of quantum gravity.
Uncertainty relations can be associated with commutators of conjugated variables
by means of the general inequality
\be
\Delta A \,\Delta B
\geq
\frac{1}{2}\left|\expec{[\hat A, \hat B]}\right|
\ .
\label{gen}
\ee
For instance, one can derive Eq.~\eqref{dxdp} from the commutator
\be
\left[\hat x, \hat p\right]
=
i\,\hslash
\left(1 + \frac{\delta_0}{\mpl^2}\, \hat p ^2\right)
\ ,
\label{gup}
\ee
for which Eq.~\eqref{gen} yields
\be
\Delta x\,\Delta p
\ge
\frac{\hslash}{2}
\left[
1+\frac{\delta_0}{\mpl^2}
\left(\Delta p^2+\expec{\hat p}^2
\right)
\right]
\ .
\label{sb}
\ee
We stress that the derivation of Eq.~\eqref{dxdp} only makes use of the algebraic structure of the commutator~\eqref{gup}
through the general inequality~\eqref{gen} and no specific representation of the physical operators $\hat{x}$ and
$\hat{p}$ (in whatsoever form) is needed.
This immediately implies that the generalised uncertainty principle~\eqref{dxdp} holds for any quantum
state, since $\expec{\hat p}^2 \geq 0$.
In particular, in the centre-of-mass frame of a scattering process, one can just consider the so-called
mirror-symmetric states satisfying $\expec{\hat p} = 0$,
and the inequality~\eqref{sb} reduces to the generalised uncertainty principle~\eqref{dxdp}.
\par
The uncertainty relation~\eqref{dxdp} readily implies the existence of a minimum (effective) length
\be
\ell=\lp\,\sqrt{\delta_0}
\ .
\ee
By comparing with Eq.~\eqref{eq:zerolength}, we obtain
\be
\delta_0
=
\frac{2}{\pi}\,
\Sigma
\ ,
\label{d0}
\ee
which represents an exact expression for the parameter of the generalised uncertainty principle emerging
from a general class of gravity theories. 
\par
For general relativity, one finds $\delta_0= {8}/{\pi}$ and $\ell\simeq 1.6\,\lp$.
Values of the deformation parameter $\delta_0$ can be obtained also from other approaches~\cite{SLV,GPLPetr} but
all available results agree in order of magnitude.
Furthermore, experimental upper bounds on $\delta_0$ exist (see Ref.~\cite{Aghababaei:2021gxe,tests} and references therein),
but they are typically too weak ($\delta_0 \lesssim 10^{36}$) to provide any useful information about the gravitational propagator.
On the other hand, the theoretical value given by Eq.~\eqref{d0} can be viewed as a (general)
lower bound.
In any case, it is hard to conceive ways of detecting direct evidence for the existence of such a vanishingly short length.
\par
The dynamical origin of the gravitational scale we discussed in this Section opens up the possibility
that much larger lengths appear in dynamical processes involving macroscopically large amounts of matter. 
For those cases, one must however face the hurdle of dealing with the nonlinearity of the gravitational
interaction beyond the weak-field approximation and perturbative methods.
%
%
%
%
%
\section{Bootstrapped Newtonian stars and black holes}
\label{sec:boot}
General relativity is a nonlinear theory already at the classical level, which is what makes it so hard to provide a quantum
description of macroscopic systems.
Starting from the Fierz-Pauli action for a spin-2 field in Minkowski spacetime, one can indeed reconstruct
the full Einstein-Hilbert action~\cite{deser}, but this bootstrap procedure is not free of
ambiguities~\cite{Padmanabhan:2004xk}.
Moreover, assuming a Minkowski background makes this approach impractical for analysing compact 
sources of astrophysical size for which the nonlinear character of gravity should be more prominent.
However, one can try and take advantage of the fact that Newtonian gravity is a well-tested approximation
as a different starting point for the bootstrap process~\cite{Casadio:2017cdv}.
\par
The bootstrapped Newtonian gravity~\cite{Casadio:2018qeh,Casadio:2019cux} is defined by considering
the well-known Newtonian Lagrangian for a spherically symmetric and static source
\be
L_{\rm N}[V]
=
-4\,\pi
\int_0^\infty
r^2 \,d r
\left[
\frac{\left(V'\right)^2}{8\,\pi\,\gn}
+\rho\,V
\right]
\ ,
\label{LagrNewt}
\ee
where $\rho=\rho(r)$ is the matter energy density.
This Lagrangian yields the Poisson equation for the Newtonian potential $V=V_{\rm N}$, that is
\be
r^{-2}\left(r^2\,V'\right)'
\equiv
\triangle V
=
4\,\pi\,\gn\,\rho
\ .
\ee
The action for bootstrapped Newtonian gravity is obtained by first adding a gravitational self-coupling
term proportional to the gravitational energy  $U_{\rm N}$ per unit volume
\be
\mathcal{J}_V
\simeq
\frac{d U_{\rm N}}{d \mathcal{V}} 
=
-\frac{\left[ V'(r) \right]^2}{2\,\pi\,\gn}
\ .
\label{JV}
\ee
One next observes that the system must be kept in equilibrium by static (and for simplicity isotropic)
pressure $p=p(r)$, which becomes increasingly large as the compactness $X \equiv {\gn\, M}/{R}$ increases.
This contribution is accounted for via a potential energy $U_p$ such that 
\be
\mathcal{J}_p
\simeq
-\frac{d U_p}{d \mathcal{V}} 
=
3\,p
\ .
\label{JP}
\ee
In the above expressions, $M$ represents an ADM-like mass and is what one would measure when studying
orbits~\cite{Casadio:2021gdf}, whereas the source of density $\rho$ and pressure $p$ will be assumed to have
finite radius $R$. 
\par
The pressure component adds to the density by effectively shifting $\rho \to \rho+3\,q_p\,p$, 
where the non-relativistic limit is obtained for the positive coupling $q_p\to 0$.
A higher-order term $\mathcal{J}_\rho=-2\,V^2$ is also added~\cite{Casadio:2017cdv}, which couples to the matter source, 
and finally leads to the total Lagrangian
\be
L[V]
&\!\!=\!\!&
L_{\rm N}[V]
-4\,\pi
\!\int_0^\infty
r^2\, d r
\!
\left[
q_V\,\mathcal{J}_V\,V
+
q_p\,\mathcal{J}_p\,V
+
q_\rho\, \mathcal{J}_\rho \left(\rho+q_p\,\mathcal{J}_p\right)
\right]
\nonumber
\\
&\!\!=\!\!& 
-4\,\pi
\!\!
\int_0^\infty
\!\!
r^2\,d r
\!\!
\left[
\frac{\left(V'\right)^2}{8\,\pi\,\gn}
\left(1-4\,q_V\, V\right)
+\left(\rho+3\,q_p\,p\right)
\!
V
\!
\left(1-2\,q_\rho\, V\right)
\!\!
\right]\!\!\!
\ ,
\label{LagrV}
\ee
where $q_V$, $q_p$ and $q_\rho$ are positive constants which track the effects of each contribution.
For instance, the Newtonian limit is recovered when all couplings vanish. 
\par
The complete Euler-Lagrange equation for the potential $V$ reads
\be
\triangle V
=
4\,\pi\,\gn\left(\rho+3\,q_p\,p\right)
\frac{1-4\,q_\rho\,V}{1-4\,q_V\,V}
+
\frac{2\,q_V\left(V'\right)^2}
{1-4\,q_V\,V}
\ ,
\label{EOMV}
\ee
which can now be considered for sources of interest. 
In particular, one must first find solutions for the outer vacuum and then boundary conditions
at $r=R$ will be used to constrain the inner solutions. 
\subsection{Outer solution}
\label{ssec:outer}
Outside the source $\rho=p=0$ and the general solution of Eq.~\eqref{EOMV} is given by
\be
V_{\rm c}
=
\frac{1}{4\,q_V}
\left[1-c_1\left(1+\frac{c_2}{r}\right)^{2/3}
\right]
\ ,
\label{sol0cc}
\ee
where $c_1$ and $c_2$ are integration constants, and it must be noted that the potential $V_{\rm c}$
cannot be shifted by an arbitrary constant because of the nonlinearity.
\begin{figure}[t]
\centering
\includegraphics[width=8cm]{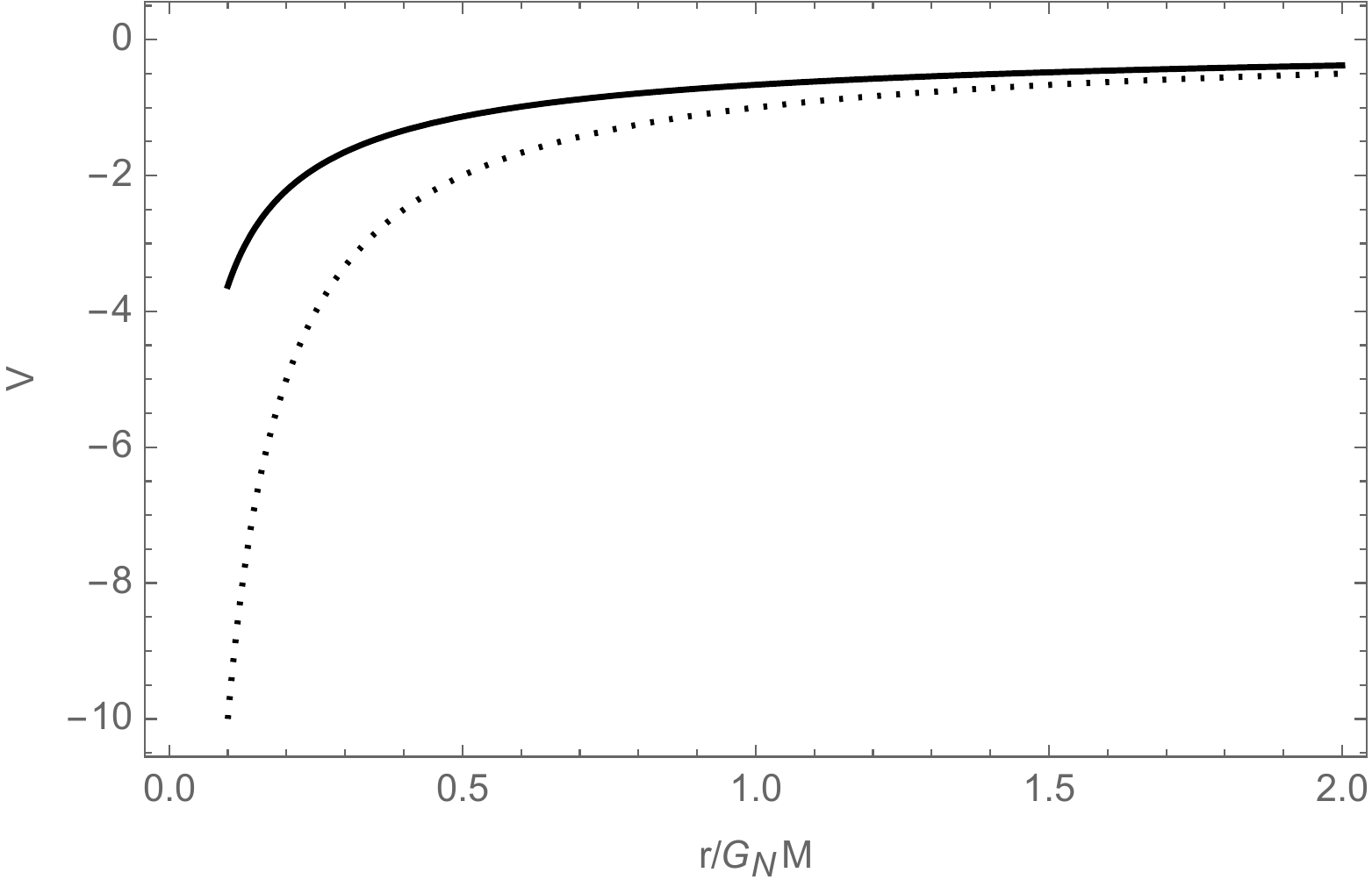}
\caption{Bootstrapped Newtonian potential~\eqref{sol0} (solid line) compared to the Newtonian
potential (dotted line).}
\label{f:Vout}
\end{figure}
\par
The Newtonian behaviour in terms of the ADM-like mass $M$ must be recovered asymptotically
for $r\to\infty$, which fixes the two integration constants and one obtains
\be
V_{\rm out}
=
\frac{1}{4\, q_V}
\left[
1-\left(1+\frac{6\,q_V \,\gn\,M}{r}\right)^{2/3}
\right]
\ .
\label{sol0}
\ee
It is easy to check that the large $r$ expansion of $V_{\rm out}$ contains the general relativity post-Newtonian
term of order $\gn^2$ of the Schwarzschild metric for $q_V=1$~\cite{Casadio:2017cdv}.
The analytic solution from Eq.~\eqref{sol0} also tracks the Newtonian potential
\be
{V}_{\rm N}
=
-\frac{\gn\,M}{r}
\ ,
\label{Vn}
\ee
for all $r>0$.
The most significant difference is that, for $r\to 0$,
\be
\frac{V_{\rm out}}{V_{\rm N}}
\sim
\left(\frac{r}{\Rh}\right)^{1/3}
\ee
and $V_{\rm out}$ diverges more slowly than $V_{\rm N}$ towards the centre (see Fig.~\ref{f:Vout}).
\subsection{Boundary conditions}
\label{ssec:bdry}
The interior solutions of Eq.~\eqref{EOMV}, $V_{\rm in}=V(0\le r<R)$, must match smoothly the outer solution
$V_{\rm out}=V(R\le r)$ from Eq.~\eqref{sol0} across the boundary $r=R$ of the source.
We therefore require that
\be
V_{\rm in}(R)
=
V_{\rm out}(R)
\equiv
V_R
=\frac{1}{4\, q_V}
\left[
1-\left(1+6\,q_V \, X\right)^{2/3}
\right] \ ,
\label{bR}
\ee
and
\be
V'_{\rm in}(R)
=
V'_{\rm out}(R)
\equiv
V'_R
=
\frac{X}
{R\left(1+6\,q_V\,X\right)^{1/3}}
\ ,
\label{dbR}
\ee
where the expressions on the right hand side are obtained by evaluating the outer potential at the boundary
and $X=\gn\,M/R$ is the outer compactness defined earlier.
Moreover, to avoid a singular centre, the classical density profiles must be finite at $r=0$,
which translates into the regularity condition 
\be
V'_{\rm in}(0)=0
\ .
\label{b0}
\ee
\par
Since Eq.~\eqref{EOMV} is a second order (ordinary) differential equation, the three constraints discussed above 
will uniquely fix the potential $V_{\rm in}$ and also determine the proper mass $M_0$ defined below. 
\subsection{Uniform sphere}
\label{ssec:homogeneous}
It is clear from Eq.~\eqref{EOMV} that finding interior solutions is going to be very complicated.
The simplest case is that of a compact homogeneous ball of matter whose density vanishes outside
the sphere of radius $r=R$,
\be
\rho
=
\rho_0
\equiv
\frac{3\, M_0}{4\,\pi\, R^3}\, 
\Theta(R-r)
\ ,
\label{HomDens}
\ee
with $\Theta$ representing the Heaviside step function. The proper mass of the object can be defined
according to the Newtonian expression
\be
M_0
=
4\,\pi
\int_0^R
r^2\,\rho(r)\, d r 
\ ,
\label{proper_mass}
\ee
and it must be emphasised that, in bootstrapped Newtonian gravity, $M_0$ is not the same as the proper mass
in general relativity (nor as the ADM mass, as we shall see below). 
\par
We must also assume a conservation equation which will further constrain the problem by relating the pressure
to the density.
We write this constraint as
\be
p'
=
-V'\left(\rho+q_p\,p\right)
\ ,
\label{eqP}
\ee
so that the it can be seen as an approximation of the Tollman-Oppenheimer-Volkoff equation of general relativity
which reduces to the Newtonian formula for $q_p=0$.
\par
For the sake of simplicity, the three couplings shall be given numerical values $q_V=q_{\rho}=1$, and for convenience
$3\,q_p = 1$. 
These assumptions reduce Eq.~\eqref{EOMV} to
\be
\triangle V
=
4\,\pi\,\gn\left(\rho+p\right)
+
\frac{2\,\left(V'\right)^2}
{1-4\,V}
\ . 
\label{EOMVs}
\ee
In the range $0\le r\le R$, the conservation Eq.~\eqref{eqP} then has the solution 
\be
\label{pressure}
p
=
\rho_0
\left[
e^{V_R-V}
-1
\right]
\ ,
\ee
where the boundary condition for the pressure $p(R)=0$ was imposed, 
$V_R$ is given in Eq.~\eqref{bR} and $\rho_0$ is defined in Eq.~\eqref{HomDens}.
Complete analytic solutions for a general value of the compactness remain too difficult to obtain
and it is convenient to separate two compactness regimes, namely $X\lesssim 1$ 
and $X\gg 1$.
\par
For low or intermediate compactness, $X \ll 1$ or $X \simeq 1$, reliable
approximate analytic solutions can be found by perturbative methods.
In particular, the potential $V_{\rm in}$ can be expanded near $r=0$ as
\be
V_{\rm s}
\simeq
V_0
+
\frac{\gn\,M_0}{2\,R^3}
\,e^{V_R-V_0}\,r^2
\ .
\label{Vs0}
\ee
where $V_0\equiv V_{\rm in}(0)<0$ and all odd powers of $r$ in the Taylor expansion must vanish
due to the regularity condition~\eqref{b0}. 
The approximate solution which matches the unique exterior~\eqref{sol0} can then be written in terms of 
the compactness $X$ as
\be
\label{Vins}
V_{\rm s}
\simeq
\frac{\left[\left(1+6\, X\right)^{1/3}-1\right]
+2\,X\left[(r/R)^2-4\right]}
{4\left(1+6\,X\right)^{1/3}}
\ ,
\label{Vintermediate}
\ee
along with the proper mass 
\be
M_0
\simeq
\frac{M\,e^{-\frac{X}{2\left(1+6\,X\right)^{1/3}}}}
{\left(1+6\,X\right)^{1/3}}
\ .
\label{M0}
\ee
This approximation of the potential can be shown to be in very good agreement with the numerical solutions
for both small and intermediate compactness, the smaller the  compactness $X$ the less $V_{\rm s}$ differs
from the numerical solution.
\par
For high compactness, $X\gg 1$, perturbative methods fail, and even the numerical analysis becomes more difficult,
because any slight error in the estimation of the proper mass $M_0=M_0(M,R)$ induces large errors in the potential profile.
In this regime, one can use comparison methods~\cite{BVPexistence,ode},
according to which upper and lower approximate bounding functions $V_\pm=V_\pm(r)$ must be somehow
found such that $E_+(r)<0$ and $E_-(r)>~0$ for the entire interval $0\le r\le R$, where
\be
E_\pm
\equiv
\triangle V_\pm
-
\frac{3\,\gn\,M_0^{\pm}(M)}{R^3}\,e^{V_R-V_\pm}
-\frac{2\left(V_\pm'\right)^2}{1-4\,V_\pm}
\ .
\label{Epm}
\ee
Mathematical theorems then ensure that the actual solution must lie in between the two bounding
functions,
\be
\label{vminvmax}
V_-<V_{\rm in}<V_+
\ .
\ee
As a further clarification, the approximate solutions $V_\pm$ are not required to have the same functional
form of the exact solution $V_{\rm in}$. 
For our case, a good starting point is the simpler equation
\be
\psi''
=
\frac{3\,\gn\,M_0}{R^3}\,e^{V_R-\psi}
\ ,
\ee 
whose relevant solutions are given by 
\be
\psi(r;M,R)
&\simeq&
\frac{X^{2/3}}{6^{1/3}}
\left(\frac{r}{R}-\frac{5}{2}\right)
\ ,
\label{psi0}
\ee 
with 
\be
\frac{M_0}{M}
&\simeq&
\left(\frac{2\,X}{3^5}\right)^{1/3}
e^{-\frac{X^{2/3}}{6^{1/3}}}
\ .
\label{M0psi}
\ee
The bounding functions for the full Eq.~\eqref{EOMV} can now be expressed as
\be
\label{VPM}
V_\pm
=
C_\pm\,\psi(r;A_\pm,B_\pm)
\ ,
\ee
where the paramaters $A_\pm$, $B_\pm$ and $C_\pm$ must be calculated using the boundary
conditions~\eqref{b0}, \eqref{bR} and~\eqref{dbR} with the additional constraint of Eq.~\eqref{Epm}.
\par
For highly compact objects with homogeneous densities, the results show a potential that is practically
linear, except in the vicinity of $r=0$ where it becomes quadratic according to Eq.~\eqref{b0}.
In the linear approximation 
\be
V_{\rm lin}
\simeq
V_R
+
V'_R\left(r-R\right)
\ ,
\label{eVlin}
\ee
where $V_R$ and $V_R'$ are given in~\eqref{bR} and \eqref{dbR}, one can also find
\be
\frac{M_0}{M}
\simeq
\frac{2\left(1+5\,X\right)}
{3\left(1+6\,X\right)^{4/3}}
\sim
\frac{1}{X^{1/3}}
\ .
\label{M0M}
\ee
\par
The bootstrapped Newtonian gravity does not account for the geometrical aspects of gravitation
and differs from general relativity in other aspects.
For example, in general relativity, black hole geometries form when matter collapses beyond the Schwarzschild radius
$\Rh=2\,\gn\,M$ and the separation between isotropic stars (with $R\gg\Rh$) and black holes
(with $R\lesssim\Rh$) is provided by the Buchdahl limit~\cite{Buchdahl:1959zz} $R>(9/8)\,\Rh$, beyond which
the necessary pressure would diverge.
\par
Bootstrapped Newtonian objects always have a finite pressure, even if their surface lies behind the horizon,
so there is no Buchdahl limit analogue.
The horizon radius $\Rh$ can still be defined using Newtonian arguments as the radius at which the escape
velocity of test particles is the same as the speed of light, or $2\,V(\rh)= -1$.
The lowest value of $X$ for which one has an horizon is then found by imposing
\be
2\, V_{\rm in}(\rh=0) 
=
-1
\ ,
\ee
which gives  $X \simeq 0.46$ if we use $V(0) = V_0$ from Eq.~\eqref{Vintermediate}. 
With increasing compactness, the radius $\rh$ grows equal to the radius $R$ of the source for
\be
2\, V_{\rm in}(\rh=R)
=
2\, V_{\rm out}(R)
=
-1
\ ,
\ee
where $V_{\rm out}(R) = V_R$ is given in Eq.~\eqref{bR}. 
The compactness for which this happens is $X \simeq 0.69$ and the horizon radius
$\rh \simeq R \simeq 1.43\, \gn\,M\simeq 3\,\gn\,M_0$.
For larger $X$, the horizon is located in the outer potential~\eqref{sol0}, and its value only depends on $M$.
\par
In summary, both the pressure and density contribute to the gravitational potential for bootstrapped Newtonian stars.
Because no equivalent to the Buchdahl limit exists, the source core can remain in equilibrium due
to a large (and finite) pressure, regardless of the compactness, with the following regimes
\be
\label{horizon}
\left\{
\begin{array}{ll}
{\rm no\ horizon}
&
{\rm for}
\ \
X
\lesssim
0.46
\\
\\
0< \rh \leq R
\simeq
1.4\,\gn\,M
\quad
&
{\rm for}
\ \
0.46 \lesssim X \leq 0.69
\\
\\
\rh
\simeq
1.4\,\gn\,M
&
{\rm for}
\ \
X \gtrsim 0.69
\ .
\end{array}
\right.
\ee
We remark that $\rh\not=\Rh$, which is a clear prediction of the bootstrapped Newtonian gravity.
\subsection{Polytropic stars and other results}
The simple model of homogenous balls was generalised to more realistic cases of stars governed
by a polytropic equation of state~\cite{Casadio:2020kbc}
\be
p=\gamma\,\rho^n
\ .
\ee
One can show numerically that the density profiles are well approximated by Gaussian functions.
Such matter distributions can then be employed in order to investigate the interplay between the compactness
$X$ of the source, the width of the density profile $\rho=\rho(r)$ and the polytropic index $n$. 
Results showed that, in the high compactness regime, bootstrapped Newtonian stars tend
to be more compact and massive than in general relativity for identical polytropic equations
of state and central densities. 
No Buchdahl limit exists in this case either and finite pressure can support a star of arbitrarily
high compactness. 
For objects with low compactness instead, the density profiles become practically identical in the two
frameworks and agree with the Newtonian approximation.
\par 
One-loop corrections to the bootstrapped Newtonian potential were computed for small
sources~\cite{Casadio:2020mch}.
On the other hand, it was shown that the potential for large homogeneous sources is not
sensitive to one-loop corrections for the matter coupling to gravity~\cite{Casadio:2019pli}.
However, the ratio between the proper mass $M_0$ and ADM mass $M$ depends on this coupling
and can be either smaller or larger than one, depending on a critical value of the coupling which,
in turn, is a function of the compactness. 
\par 
Another case worth mentioning is the investigation of binary mergers~\cite{Casadio:2022gbv},
where several constraints resulting from the difference between the ADM mass $M$
and the proper mass $M_0$ can be imposed.
These constraints change depending on the types of stars that merge and resulting final object.
The more interesting finding regards the GW150914 signal measured by LIGO:
contrary to some other claims found in the literature, LIGO's findings do not violate
the mass gap in bootstrapped Newtonian gravity, but typical stellar black hole masses easily
fit the experimental data~\cite{LIGOScientific:2018mvr}. 
\par
This classical picture can change significantly if one considers the quantum nature of matter
and, consequently, of gravity.
A first indication that the existence of a dynamical length similar to the one discussed in
Section~\ref{sec:minlength} leads to an upper bound on the compactness was discussed in
Ref.~\cite{Casadio:2020ueb} within the bootstrapped Newtonian picture.
However, it is indeed simpler to reach that conclusion by starting from full general relativity, as we are going to see next.
%
%
%
%
%
%
\section{Quantum matter core}
\label{sec:core}
Semiclassical models of gravitational collapse generically predict a bounce at a minimum
radius~\cite{frolov,Casadio:1998yr,hajicek,hk},
but the nonlinearity of the Einstein equations makes it impossible to study realistic models analytically, which
renders the problem of a quantum description intractable.
For this reason, one can only study very simplified models obtained by forcing a strong symmetry and unphysical equation of state
for the collapsing matter.
A paramount example is given by the gravitational collapse of a ball of dust (matter with no other interaction but gravity)
originally investigated by Oppenheimer and collaborators~\cite{OS}.
A similarly simple case is given by a shell of matter collapsing under its weight or towards a central source.
\par
In fact, these systems are simple enough to allow for a canonical analysis~\cite{kiefer} of the effective action obtained by restricting
the Einstein-Hilbert action to metrics satisfying the assumed symmetry properties.
Some key features are more readily obtained with the approach of Refs.~\cite{Casadio:2021cbv,Casadio:2022pla,Casadio:2022epj},
in which the geodesic equation for the areal radius of the ball provides an effective quantum mechanical description similar to the
usual quantum model of the hydrogen atom obtained by quantising the electron's position.
This approach straightforwardly leads to the existence of a discrete spectrum of bound states,
with the ground state characterised by a macroscopically large surface area quantised according to Bekenstein's
area law~\cite{bekenstein}.
It is important to remark from the very beginning that the areal radius of the ball is not a fundamental degree of freedom for the matter
in the collapsing core, which should instead be described by quantum excitations of fields in the Standard Model of particle physics.
Although these fundamental degrees of freedom are neglected for the purpose of defining a tractable mathematical
problem, their existence must be encoded in a suitable entropy that can be computed from the states
of the effective quantum mechanical theory.
\subsection{General relativistic discrete spectrum}
Let us consider a perfectly isotropic ball of dust with total ADM mass $M$ and areal radius $R=R(\tau)$,
where $\tau$ is the proper time measured by a clock comoving with the dust. 
Dust particles inside this collapsing ball will follow radial geodesics $r=r(\tau)$ in the Schwarzschild space-time
metric
\be
\d s^{2}
=
-\left(1-\frac{2\,\gn\,M_\mu}{r}\right)
\d t^{2}
+
\left(1-\frac{2\,\gn\,M_\mu}{r}\right)^{-1}
\d r^{2}
+r^{2}\,\d\Omega^2
\ ,
\label{schw}
\ee
where $M_\mu=M_\mu(r)$ is the (constant) fraction of ADM mass inside the sphere of radius $r=r(\tau)$.
In particular, we can consider the outermost (thin) layer of (average) radius $r=R(\tau)$ and mass $\mu=\epsilon\,M$,
with $0<\epsilon<1$.
The evolution of $R$ is then determined by the radial equation~\footnote{One should notice that Eq.~\eqref{geod-general}
holds both in a black hole and in a white hole background.
Classically, geodesics falling into the black hole singularity cannot be continued into geodesics emerging from the 
white hole singularity.
However, quantum physics bypasses this limitation and one finds bouncing solutions which do precisely that
in the semiclassical analysis of the model.}
\be
\label{geod-general}
\frac{E_{\mu}^{2}}{\mu^{2}}
-
\dot R^{2}
+
\frac{2\,\gn\,M_\mu}{R}
=
1
\ ,
\ee
where $M_\mu=(1-\epsilon)\,M$ and $E_\mu$ is the conserved momentum conjugated to
$t=t(\tau)$.~\footnote{Of course, the conserved angular momentum vanishes for purely radial motion.
For a (slowly and rigidly) rotating ball see Ref.~\cite{Casadio:2022epj}.}
Eq.~\eqref{geod-general} can be written as
\be
\label{geod-part}
H
\equiv
\frac{P^{2}}{2\,\epsilon\, M}
-\frac{\epsilon\left(1-\epsilon\right)\gn\,M^{2}}{R}
=
\frac{\epsilon\, M}{2}\left(\frac{E_\mu^{2}}{\epsilon^2\,M^{2}}-1\right)
\equiv
\mathcal E
\ ,
\ee
where $P=\mu\,\dot R$ is the momentum conjugated to $R=R(\tau)$.
\par
We can next apply the canonical quantisation prescription $\hat{P}=-i\,\hslash\,\partial_R$,
which allows us to write Eq.~\eqref{geod-part} as the time-independent Schr\"odinger equation
\be
\hat{H}\,\Psi_{\bar n}
=
\left[
-
\frac{\hslash^{2}}{2\,\epsilon\, M}
\left(
\frac{\d^2}{\d R^2}
+
\frac{2}{R}\,
\frac{\d}{\d R}
\right)
-
\frac{\epsilon\left(1-\epsilon\right)\gn\,M^{2}}{R}
\right]
\Psi_{\bar n}
=
\mathcal E_{\bar n}\,
\Psi_n
\ .
\label{HpsiEpsi}
\ee
Since this is formally the same as the equation for the $s$-states of the hydrogen atom,
solutions are given by the  eigenfunctions
\be
\label{radial-wavefunction}
\Psi_{\bar{n}}
=
\sqrt{\frac{\epsilon^{6}\left(1-\epsilon\right)^{3}M^{9}}{\pi\,\lp^{3}\,\mpl^{9}\,\bar{n}^{5}}}\,
\exp\!\left(-\frac{\epsilon^{2}\,(1-\epsilon)\,M^{3}\,R}{\bar{n}\,\mpl^{3}\,\lp}\right)
L_{\bar{n}-1}^{1}\!\!
\left(\frac{2\,\epsilon^{2}\,(1-\epsilon)\,M^{3}\,R}{\bar{n}\,\mpl^{3}\,\lp}\right)
\!\!\!
\ ,
\quad
\ee
where $L_{{\bar n}-1}^1$ are Laguerre polynomials with ${\bar n}=1,2\,\ldots$, corresponding to the eigenvalues 
\be
\mathcal E_{\bar n}
=
-
\frac{\epsilon^3\,(1-\epsilon)^2\,M^5}{2\,\mpl^4\,\bar n^2}
\ .
\ee
The normalisation is defined in the scalar product which makes $\hat H$ Hermitian on the above spectrum, that is
\be
\pro{\bar n}{\bar n'}
=
4\,\pi\,\int_0^\infty
R^2\,\Psi_{\bar n}^*(R)\,\Psi_{\bar n'}(R)\,
\d R
=
\delta_{\bar n \bar n'}
\ ,
\ee
and the expectation value of the areal radius is thus given by
\be
\bar R_{\bar n}
\equiv
\bra{\bar n} \hat R \ket{\bar n}
=
\frac{3\,\mpl^3\,\lp\,\bar n^2}{2\,\epsilon^2\,(1-\epsilon)\,M^3}
\ .
\ee
So far the quantum picture is the same that one would have in Newtonian physics, with the ground state $\bar n=1$
having a width $\bar R_1\sim \lp\,(\mpl/M)^3$ and energy $\mathcal E_1\sim -M\,(M/\mpl)^4$.
This state is practically indistinguishable from a point-like singularity for any macroscopic black hole of mass $M\gg\mpl$.
\par
In fact, the only general relativistic feature that the model retains is given by the relation for $\mathcal E=\mathcal E(E_\mu)$
in Eq.~\eqref{geod-part}.
On assuming that $E_\mu$ is real for the allowed quantum states, we obtain $E_\mu^2\ge 0$,
which yields the bound 
\be
\bar n
\ge
N_M
\equiv
\epsilon\,(1-\epsilon)
\left(\frac{M}{\mpl}\right)^2
\ .
\label{N_M}
\ee
The fundamental state of the outer layer saturates the inequality \eqref{N_M}, which is again
reminiscent of Bekenstein's area law~\cite{bekenstein}.~\footnote{It is interesting to recall that a similar bound
was also found for stable configurations of boson stars~\cite{Kaup:1968zz,Ruffini:1969qy}.}
This result implies that the probability to find the ball of dust near $r=0$ is practically zero for $M\gg\mpl$
and the singularity is avoided.
Classically, there are no reasons for a ball of dust evolving solely under the gravitational force to stop contracting,
but quantum mechanics instead shows that, in order to have a well-defined energy spectrum, the proper ground state compatible
with general relativity has a width
\be
R_M
\equiv
\bar R_{N_M}
=
\frac{3}{4}\,(1-\epsilon)\,\Rh
\ ,
\label{R_M}
\ee
where $\Rh=2\,\gn\,M$ is again the classical Schwarzschild radius of the ball.
\par
From the wavefunction $\Psi=\Psi(R)$, we can determine the probability that the ball be inside its own
gravitational radius,
\be
P(R\le \Rh)
\equiv
\int_0^{\Rh}
\mathcal P(R)\,\d R
=
4\,\pi
\int_0^{\Rh}
|\Psi(R)|^2\,
R^2\,\d R
\ ,
\ee
which can be viewed as the probability that the dust ball is a black hole (when the mass $M$ is
treated as a fixed parameter~\footnote{An alternative viewpoint is considered in Refs.~\cite{qbh2,Casadio:2013uga},
where the mass $M$ is quantised.}).
For the ground state and values of $M\gg \mpl$, the probability density $\mathcal P=\mathcal P_{N_M}(R)$ 
narrowly peaks around a value of $R=R_M$ below $\Rh$ and one thus finds 
\be
P(R\le \Rh)\simeq 1
\ ,
\label{Pbh}
\ee
which confirms that the ground state is (very liekly) a black hole.
\par
We further notice that the uncertainty in the areal radius is given by
\be
\frac{\Delta R_{\bar n}}{\bar R_{\bar n}}
\equiv
\frac{\sqrt{\bra{\bar n}\hat R^2\ket{\bar n}-\bar R_n^2}}{\bar R_n}
=
\frac{\sqrt{\bar n^2+2}}{3\,\bar n}
\ ,
\ee
which approaches a minimum of $1/3$ for $\bar n\to \infty$.
This supports the conclusion that the matter core of a black hole is indeed in a relatively ``fuzzy'' quantum state
(whereas excited states are more and more classical).
Since $0<\epsilon<1$, one can conclude that $R_M<\Rh$ and a finite number of states~\eqref{radial-wavefunction}
with $\bar n=N_M+n$ and $n=0,1,\ldots$ will exist inside the horizon.
\par
Although we (formally) know the analytical expression for the spectrum of the quantum dust ball,
it is technically very difficult to perform explicit calculations for $M\gg\mpl$.
For instance, for $M=M_\odot\simeq 10^{30}\,$kg, one finds that the ground state has $N_M\sim 10^{76}$
nodes, which makes it impossible to handle the wavefuntion~\eqref{radial-wavefunction}, either analytically or numerically.
\subsection{Entropy and thermodynamics}
\label{S:entropy}
The bound on the compactness following from Eq.~\eqref{N_M} is caused by the nonlinearity
of general relativity and qualitatively agrees with the results obtained by adding a gravitational self-interaction
term to the Newtonian theory described in the previous Section~\ref{sec:boot}.
It also agrees with those results following from the quantum description of the gravitational radius and
black hole horizon~\cite{qbh2,Casadio:2013uga}.
In particular, the latter approach leads to very similar conclusions when the
self-gravitating object is described by an extended many-body system with a very large occupation
number of order $N_M\sim M^2/\mpl^2$ in its ground state~\cite{Casadio:2015bna}.
The first excited modes could then be populated thermally and reproduce
the Hawking radiation~\cite{hawking}.
\par
In fact, we notice that, for $\bar n\gtrsim N_M$, the quantum of the Hamiltonian $H$ in Eq.~\eqref{HpsiEpsi}
is given by
\be
\delta H
\equiv
\left|\mathcal E_{n+1}
-
\mathcal E_{n}
\right|
\simeq
\mpl\,\frac{\mpl}{M}
\ ,
\label{dH}
\ee
so that $\delta H\ll \mpl$ for a macroscopic object of mass $M\gg \mpl$.
This is the quantum of horizon area $\delta M\simeq \delta H$ predicted by Bekenstein and the typical energy of
Hawking quanta.
Furthermore, it appears that the proper source ``energy'' $\delta E \equiv \left|E_{n+1} -E_{n}\right| \simeq \mpl$
is naturally quantised in units of the fundamental Planck mass $\mpl$, but this quantum is redshifted down
to the much smaller $\delta H$ measured by outer observers.
It is intriguing that Eq.~\eqref{N_M} allows for recovering the fundamental scaling relations
of the corpuscular description of black holes~\cite{DvaliGomez,Dvali:2012rt} in which the Hawking evaporation process
is described as the depletion of the quantum state of gravity~\cite{DvaliGomez,Dvali:2012rt}.
\par
In this perspective, the wavefunction $\Psi_{N_M}$ appears as the non-perturbative ground state
for self-gravitating macroscopic objects of mass $M$ and should thus be the closest possible to
a classical black hole.
This result is consistent with the fact that a static gravitational field must be fully determined by the
state of the source~\cite{qbh2, Casadio:2013uga}.
It can further be interpreted as the fact that the quantum state of a macroscopic self-gravitating
system of mass $M$ is very far from the vacuum $\Psi_0$ of quantum gravity, for which 
$N_{M}=0$.
The number $N_M\sim M^2/\mpl^2$ hence provides a quantitative measure
for this ``distance'' from the vacuum in the Hilbert space of quantum gravity states.
The appearance of this ground state, in turn, can be interpreted as a form of {\em classicalization\/}
ensuring the ultraviolet self-completeness of gravity~\cite{Dvali:2010ns,classicalisation}, because states with spatial
momenta much larger than $\hslash/R_M\sim \hslash/\Rh$ have $\bar n<N_M$ and cannot be populated
at energy scales of the order of the mass $M$.
\par
Given that the collapsing objects should contain a very large number of matter field excitations, whose states
are completely neglected, two different types of information entropy were employed
to probe this scenario in Ref.~\cite{Casadio:2022pla}, namely the configurational entropy
and the continuous limit of Shannon's entropy.
Both can only be computed numerically and for relatively small principal quantum numbers $\bar n=N_M+n$,
for which they increase with $M$, in agreement with the intuition that,
the larger the mass the more constituents in the dust core, with consequently larger uncertainty in its microstates.
Excited states (with $\bar n>N_M$) display (much) higher information entropy than the ground state
for the same mass $M$, which also agrees with the intuition that excited states are spatially less localised and can
be realised by larger numbers of microstates. 
These results are in particular compatible with the classical dynamics of a ball of dust, which is necessarily going
to collapse under its own weight without loss of energy encoded by the ADM mass $M$.
One can then reproduce the time evolution of the areal radius $R(\tau)=\bar R_{\bar n}$ with decreasing values of
$\bar n=N_M+n$ in correspondence with increasing proper time $\tau$.
The existence of a ground state halts this process at a finite macroscopic size $\bar R=R_M$, after which the ball
can only shrink further by losing energy $M$, so that $N_M$ decreases.
This behaviour is qualitatively very different from the hydrogen atom, in which an electron bound to the nucleus
will jump into smaller quantum states necessarily by emitting energy in the form of radiation.
The dust ball instead collapses without emitting (gravitational) energy, because of the strict spherical symmetry,
until it reaches a minimum (quantum) size (although one expects that a realistic collapse
is instead always associated with the emission of energy).
Finally, both entropies are smaller for larger fractions $\epsilon=\mu/M$ of dust in the outermost layer.
Since larger values of the entropy should correspond to more unstable configurations, this result seems to favour the
accumulation of matter in the outermost layer, due to quantum pressure.
%
%
%
%
\section{Coherent state for the ``hairy'' geometry}
\label{sec:coherent}
The main conclusion of the previous Section was that matter at the end of the gravitational collapse
forms an extended quantum core of width $R_M\lesssim \Rh=2\,\gn\,M$.
It is conceivable that this feature reflects in the quantum state of the outer geometry, by showing some
sort of deviation (or quantum ``hair'') from the classical general relativity solutions, albeit without changing the
fundamental dynamics of general relativity.
In fact, while the existence of particular quantum states can radically change the description of a system,
at the same time it can leave the underlying fundamental dynamics unaffected.
One must also bare in mind that, for an astrophysical black hole, the  resulting effectively classical
description should be consistent with the phenomenology of spacetime
that is already confirmed experimentally~\cite{Goddi:2019ams,LIGOScientific:2018mvr}.
\par
In the vast landscape of quantum models of black holes, the corpuscular picture~\cite{DvaliGomez,Dvali:2012rt}
is among those in which geometry emerges at some macroscopic scale from the microscopic quantum theory of
gravitons~\cite{feynman,deser}.
In this model, soft (virtual) gravitons are marginally bound in the potential well that they create and
are characterised by a Compton-de~Broglie wavelength $\lambda_{\rm G}$ equal to
the horizon radius $\Rh$ of the black hole.
One can easily infer that the energy scale of the gravitons is on the order of
$\epsilon_{\rm G}\sim \hslash/\lambda_{\rm G}$. 
For a total number $N_{\rm G}$ of gravitons, the total mass is $M\simeq N_{\rm G}\,\epsilon_{\rm G}$,
and one can infer the scaling relations $N_{\rm G}\sim{M^2}/{\mpl^2}\sim{\Rh^2}/{\lp^2}$.
The nonlinearity of the gravitational interaction is apparent from the (negative) gravitational energy
for an object of mass $M$ contained within a sphere of radius $R$,
that is $U_{\rm N}\sim M\, V_{\rm N}(R)$, with ${V}_{\rm N}$ given in Eq.~\eqref{Vn}.
Most gravitons in this black hole model should have roughly the same wavelength $\lambda_{\rm G}$
and individual binding energy
\be
\epsilon_{\rm G}
\sim
\frac{U_{\rm N}}{N_{\rm G}}
\sim
-\frac{\lp\,\mpl}{R} 
\ ,
\ee
from which the Compton-de~Broglie length $\lambda_{\rm G}\sim R$.
The graviton self-interaction energy then reproduces the positive post-Newtonian energy
\be
U_{\rm GG}
\sim
N_{\rm G} \, \epsilon_{\rm G} \, V_{\rm N}(R)
\sim
\frac{\gn^2\,M^3}{R^2}
\ .
\label{Ugg}
\ee
Assuming that the gravitons trapped inside the black hole are marginally bound, that is
$U_{\rm N}+U_{\rm GG}\simeq 0$, one finally finds
$\lambda_{\rm G}\sim R\simeq \Rh$~\cite{Casadio:2016zpl,Casadio:2017cdv}.
Like for all bound states in quantum physics, modes of arbitrarily small wavelengths do not appear and, as a result, 
the classical central singularity is not realised.
This is another form of {\em classicalization\/} of gravity~\cite{Dvali:2010ns,classicalisation} and,
as we recalled in previous Sections, this feature can also be explained from the quantum
nature of the source.
\par
Modelling black holes so simply as quantum states of gravitons with similar wavelengths $\lambda_{\rm G}$
is quite restrictive, since the inhomogeneous gravitational field in the space outside the object cannot be reproduced,
not even in the Newtonian approximation.
General relativity is a metric theory and without going as far as a full quantum theory of gravity, 
this problem can be approached in a simpler way for static and spherically symmetric
sources~\cite{Casadio:2016zpl,Giusti:2021shf,Casadio:2022ndh}.
More precisely, the quantum description can be limited to the metric function~\eqref{Vn} that appears in the
Schwarzschild metric
\be
\d s^2
=
-\left(1+2\,V_{\rm N}\right)\d t^2
+
\left(1+2\,V_{\rm N}\right)^{-1}\d r^2
+
r^2\,\d\Omega^2
\ .
\label{gSch}
\ee
We remind that $V_{\rm N}$ represents the potential in the radial geodesic equation~\eqref{geod-part}
from which we derived the discrete spectrum in Section~\ref{sec:core}.
However, while $r$ in Eq.~\eqref{gSch} is the areal radius, the (post-)Newtonian
expressions above and in Section~\ref{sec:boot} should be given in terms of the harmonic
radius~\cite{weinberg,Casadio:2021cbv,Casadio:2021gdf}.
\par
We next assume the quantum vacuum $\ket{0}$ to be a state of the universe with no excited modes
(of matter or gravity).
The Minkowski metric $\eta_{\mu\nu}$ can be associated to this absolute vacuum,
because it is used to describe linearised gravity and for the definition of matter and gravitational
excitations under these circumstances.
Small matter sources with energy $M\ll\mpl$ should be described accurately in this regime.
Also, the Newtonian potential should be recovered from tree-level graviton exchanges in the
non-relativistic limit.
The potential $V_{\rm N}$ therefore emerges from the (non-propagating) longitudinal polarisation
of virtual gravitons in this limit~\cite{feynman}.
For sources with masses much larger than the Planck mass $M\gg\mpl$ the complete
classical dynamics can be in principle reconstructed~\cite{deser,Padmanabhan:2004xk}.
However, calculating the proper quantum state starting from the excitations
of the linearised theory looks impossible in a highly nonlinear regime.
This problem is usually bypassed (although not solved) in the effective field theory approach by
assuming the existence of a classical background geometry which replaces $\eta_{\mu\nu}$
with $g_{\mu\nu}$ and which should be a solution to the classical Einstein equations.
\par
This assumption is necessary if one wants to compare the theory with experimental data
available in the weak-field regime, where the Einstein equations well describe
the dynamics at macroscopic scales.
We shall therefore assume that the relevant quantum states of gravity for static and spherically
symmetric black holes effectively reproduce the Schwarzschild geometry~\eqref{gSch}.
With this in mind, what is proposed next can be seen as quantising the longitudinal
mode of the gravitational field using coherent states
(which have the property of minimising the quantum uncertainty) in a suitable
Fock space that is built starting from the Minkowski background
vacuum~\cite{Casadio:2021eio,Mueck:2013mha,Bose:2021ytn,Berezhiani:2021zst}.
\subsection{Quantum coherent states for classical static configurations}
\label{coherent}
Given the previous discussion, a generic static metric function $V=V(r)$ can be conveniently described
as the mean field of the coherent state of a free massless scalar field~\cite{Casadio:2017cdv}.
The dimensionless potential $V$ must first be rescaled to arrive at a canonically normalised real scalar field
$\Phi =\gn\,V=\sqrt{{\mpl}/{\lp}}\, V$.  The next step is to quantise $\Phi$ as a massless field which must satisfy the
wave equation
\be
\left[
-\frac{\partial^2}{\partial t^2}
+
\frac{1}{r^2}\,\frac{\partial}{\partial r}
\left(r^2\,\frac{\partial}{\partial r}\right)
\right]
\Phi(t,r)
\equiv
\left(-\partial_t^2+\triangle\right)
\Phi
=
0
\ ,
\label{KG}
\ee
whose solutions can be expressed as
\be
u_{k}(t,r) = e^{-i\,k\,t}\,j_0(k\,r)
\ ,
\label{u_k}
\ee
where $k>0$ and $j_0={\sin(k\,r)}/{k\,r}$ represents the spherical Bessel function 
satisfying the orthogonality condition 
\be
4\,\pi
\int_0^\infty
r^2\,\dd r\,
j_0(k\,r)\,j_0(p\,r)
=
\frac{2\,\pi^2}{k^2}\,
\delta(k-p)
\ .
\ee
The quantum field operator, respectively its conjugate momentum are of course given by
\be
\label{Phi}
\hat{\Phi}(t,r)
\!\!&=&\!\! 
\int_0^\infty
\frac{k^2\,\dd k}{2\,\pi^{2}}\,
\sqrt{\frac{\hslash}{2\,k}}
\left[
\a_{k}\,
u_k(t,r)
+
\ac_{k}\, 
u^*_k(t,r)
\right] \ ,
\\
\hat{\Pi}(t,r) 
\!\!&=&\!\!
i\int_0^\infty
\frac{k^2\,\dd k}{2\,\pi^{2}}\,
\sqrt{\frac{\hslash\,k}{2}}
\left[
\a_{k}\,
u_k(t,r)
-
\ac_{k}\, 
u^*_k(t,r)
\right]
\ . 
\ee
The two operators must satisfy the equal time commutation relations,
\be
\left[\hat{\Phi}(t,r)
,\hat{\Pi}(t,s)\right] 
=
\frac{i\,\hslash}{4\,\pi\,r^2}\,
\delta(r-s)
\ ,
\ee
so that the creation and annihilation operators satisfy the commutation rules
\be
\left[\a_{k},\ac_{p}\right]
=
\frac{2\,\pi^2}{k^2}\,
\delta(k-p)
\ .
\ee
As usual, the Fock space is constructed from the vacuum defined as $\a_{k}\ket{0}=0$ for all $k>0$.
The background we have chosen to correspond to the state $\ket{0}$ is the flat Minkowski metric which defines
the d'Alembert operator in Eq.~\eqref{KG}, precisely because this vacuum $\ket{0}$ is supposed to describe
a spacetime that is completely empty. 
One could in fact have started from a space of any dimension, since without events, the notions of
spatial dimension and time are empty (and the entire spacetime manifold cannot be distinguished from
a point).
Knowing that the system we eventually want to describe is static and three-dimensional, we explicitly
introduced three spatial dimensions and one time from the onset. 
\par
The classical configurations of $\Phi$ which can be obtained in the quantum theory
must be associated with certain states in this Fock space.
In particular, coherent states 
\be
\a_{k} \ket{g} = g_{k}\,e^{i\,\gamma_{k}(t)} \ket{g}
\ee
represent a good option provided the expectation values of $\hat\Phi$ reproduce the classical potential,
that is
\be
\gn\,
\bra{g}\hat{\Phi}(t,r)\ket{g}
=
V(r)
\ .
\label{expecphi}
\ee
From Eq.~\eqref{Phi} one finds
\be
\bra{g}\hat{\Phi}(t,r)\ket{g}
=
\int_0^\infty
\frac{k^2\,\dd k}{2\,\pi^2}\,
\sqrt{\frac{2\,\lp\,\mpl}{k}}\,
g_k\,
\cos[\gamma_k(t)-k\,t]\,
j_0(k\,r)
\ .
\ee
By writing the potential as
\be
V
=
\int_0^\infty
\frac{k^2\,\dd k}{2\,\pi^2}\,
\tilde V(k)\,
j_0(k\,r)
\ ,
\label{Vk}
\ee
the condition of perfect staticity is obtained for $\gamma_k=k\,t$ and
\be
g_k
=
\sqrt{\frac{k}{2}}\,
\frac{\tilde V(k)}{\lp}
\ .
\label{gkVk}
\ee
The above choice of phases is a rather utilitarian way to eliminate the time dependence from
the normal modes~\eqref{u_k} and recover static configurations.
From a physical perspective, no real system is perfectly static and this choice is a good
approximation as long as the actual time evolution occurs on time scales much longer
than $\Delta t\sim k^{-1}$.
\par
Putting together these results, the coherent state becomes
\be
\ket{g}
=
e^{-N_{\rm G}/2}\,
\exp\left\{
\int_0^\infty
\frac{k^2\,\dd k}{2\,\pi^2}\,
g_k\,
\ac_k
\right\}
\ket{0}
\ ,
\label{gstate}
\ee
with
\be
\label{NGN}
\Ng
=
\int_0^{\infty} 
\frac{k^2\,\dd k}{2\,\pi^2}\, 
g_k ^2
\ee
formally representing the total occupation number.
A more explicative meaning is perhaps that the value of $\Ng$ measures the ``distance''
in the Fock space between $\ket{g}$ and the vacuum $\ket{0}$, the latter corresponding to $\Ng=0$.
Another quantity of interest is given by
\be
\label{EN}
\expec{k}
=
\int_0^{\infty} 
\frac{k^2\,\dd k}{2\,\pi ^2} 
\,k\,g_k ^2
\ , 
\ee
from which the ``average'' wavelength $\lambda_{\rm G}=\Ng/\expec{k}$ is obtained.
\subsection{Quantum Schwarzschild black holes}
\label{S:QBH}
The Schwarzschild metric~\eqref{gSch} is completely defined by $V_{\rm N}=\sqrt{\gn}\,\Phi$ in Eq.~\eqref{Vn}, and 
all the necessary quantities can be calculated from the occupation numbers $|g_k|^2$ of the modes $u_k$.
Eq.~\eqref{Vk} can be easily inverted to obtain 
\be
\tilde{V}_{\rm N}
=
-4\,\pi\,\gn\,\frac{M}{k^2}
\ee
and the coefficients
\be
g_k
=
-\frac{4\,\pi\,M}{\sqrt{2\,k^3}\,\mpl}
\ .
\label{gkN}
\ee
The total occupation number for such a coherent state is then formally given by
\be
N_{\rm G}
=
4\,\frac{M^2}{\mpl^2}
\int_0^\infty
\frac{\dd k}{k}
\ ,
\ee
which suffers of logarithmic divergences both in the infrared and in the ultraviolet.
Likewise, we can compute the average wavenumber
\be
\expec{k}
=
4\,\frac{M^2}{\mpl^2}
\int_0^\infty
\dd k
\ ,
\ee
which only diverges (linearly) in the ultraviolet.
\par
The ultraviolet divergences appear in the above expression because the Schwarz\-schild geometry~\eqref{gSch}
was assumed to hold for all values of $r>0$, corresponding to a vanishing size of the central matter source.
On the other hand, a diverging $N_{\rm G}$ implies that the coherent state $\ket{g}$ is not normalisable and 
one therefore arrives at the strong conclusion that the classical vacuum Schwarzschild geometry cannot be
realised in quantum gravity~\cite{Casadio:2021eio}.
Equivalently, the very existence of a proper quantum state $\ket{g}$ requires that the coefficients $g_k$ depart
from their classical expression~\eqref{gkN} for $k\to 0 $ and $k\to\infty$.
\par
An obvious solution are sources with a finite size $R$, such as the quantum dust cores investigated
in Section~\ref{sec:core}, or even with regular density profiles like the ones in Section~\ref{sec:boot}. 
Either case would lead to occupation numbers $|g_k|^2$ which are sufficiently suppressed for $k\gg R^{-1}$
to remove the ultraviolet divergences. 
(In the case of black holes, $R$ must be smaller than $\Rh$.)
A detailed knowledge of the (effective) energy-momentum tensor of these sources would be
necessary to compute explicitly the coefficients $g_k$, but this information is not yet available for the quantum
dust core.
For the sole purpose of removing the ultraviolet divergences, it is however enough to impose a cut-off
$k_{\rm UV}\sim 1/R$, where we can eventually set $R\simeq R_M$ in Eq.~\eqref{R_M} for a formed
black hole.
\par
In the infrared regime, we can impose a cut-off  $k_{\rm IR} = 1/\Rinf$ which accounts for
the finite life-time $\tau\sim \Rinf$ of a realistic source~\cite{Casadio:2017cdv}.
For $R_\infty$ one can also consider as an upper bound the size of the observable
Universe~\cite{Giusti:2021shf}.
However, this does not affect the results very much because modes with wavelength $k^{-1}$
that are orders of magnitude larger than the horizon radius $\Rh$ do not impact
the expectation value in Eq.~\eqref{expecphi} significantly~\cite{Casadio:2020ueb,DvaliSoliton}.
The occupation number now becomes 
\be
N_{\rm G}
=
4\,\frac{M^2}{\mpl^2}
\int_{k_{\rm IR}}^{k_{\rm UV}}
\frac{\dd k}{k}
=
4\,\frac{M^2}{\mpl^2}\,
\ln\left(\frac{R_\infty}{R}\right)
\label{cNg}
\ee
and the average wavenumber
\be
\expec{k}
=
4\,\frac{M^2}{\mpl^2}
\int_{k_{\rm IR}}^{k_{\rm UV}}
\dd k
=
4\,\frac{M^2}{\mpl^2}
\left(
\frac{1}{R}
-
\frac{1}{R_\infty}
\right)
\ .
\label{ckg}
\ee
The expression for $N_{\rm G}$ shows the same leading dependence on $M^2$ as the
corpuscular scaling relation and Bekenstein's area law.
The other essential result is that
\be
\lambda_{\rm G}
=
\frac{N_{\rm G}}{\expec{k}}
\sim 
\lp\,\frac{M}{\mpl}
\ ,
\ee
for $R\simeq R_M\sim\Rh$.
One can further speculate that, because of the logarithm in Eq.~\eqref{cNg},
the number $N_{\rm G}$ will grow during the collapse,
when the radius $R$ of the matter source decreases, until the maximum
is reached for $R\simeq R_M$ corresponding to the core ground state.
\par
For the model to be realistic and comparable with experimental bounds, the mean field must still accurately
reproduce the classical $V_{\rm N}$, at least in the region outside the black hole horizon.
Therefore, the coherent state $\ket{g_{\rm BH}}$ which represents astrophysical black holes should give
\be
\gn
\bra{g_{\rm BH}}\hat{\Phi}(t,r)\ket{g_{\rm BH}}
\simeq
V_{\rm N}(r)
\qquad
{\rm for}
\
r\gtrsim \Rh
\ ,
\label{QCo}
\ee
where the approximate equality was used to suggest that this is up to the experimental precision. 
Because of the previous discussion, the mean field in the left hand side of Eq.~\eqref{QCo} will
necessarily involve deviations from the classical function in the right hand side.
In fact, one can easily compute the effective quantum potential
\be
V_{\rm QN}
\simeq
\int_{k_{\rm IR}}^{k_{\rm UV}}
\frac{k^2\,\dd k}{2\,\pi^2}\,
\tilde V_{\rm N}(k)\,j_0(k\,r)
\simeq
-
\frac{2\,\lp\,M}{\pi\,\mpl\,r}
\int^{r/R}_{0}
\dd z\,
\frac{\sin z}{z}
\ ,
\label{Vqq}
\ee
where, in the second line, the substitution $z=k\,r$ was made and for the infrared cut-off the
limit $k_{\rm IR}=1/R_\infty\to 0$ was taken.
Finally, one obtains
\be
V_{\rm QN}
\simeq
V_{\rm N}
\left\{
1
-\left[1-
\frac{2}{\pi}\,{\rm Si}\left(\frac{r}{R}\right)
\right]
\right\}
\ ,
\label{Vq}
\ee
where ${\rm Si}$ represents the sine integral function.
\subsection{Geometry}
\label{A:geo}
\begin{figure}[t]
\centering
\includegraphics[width=8cm]{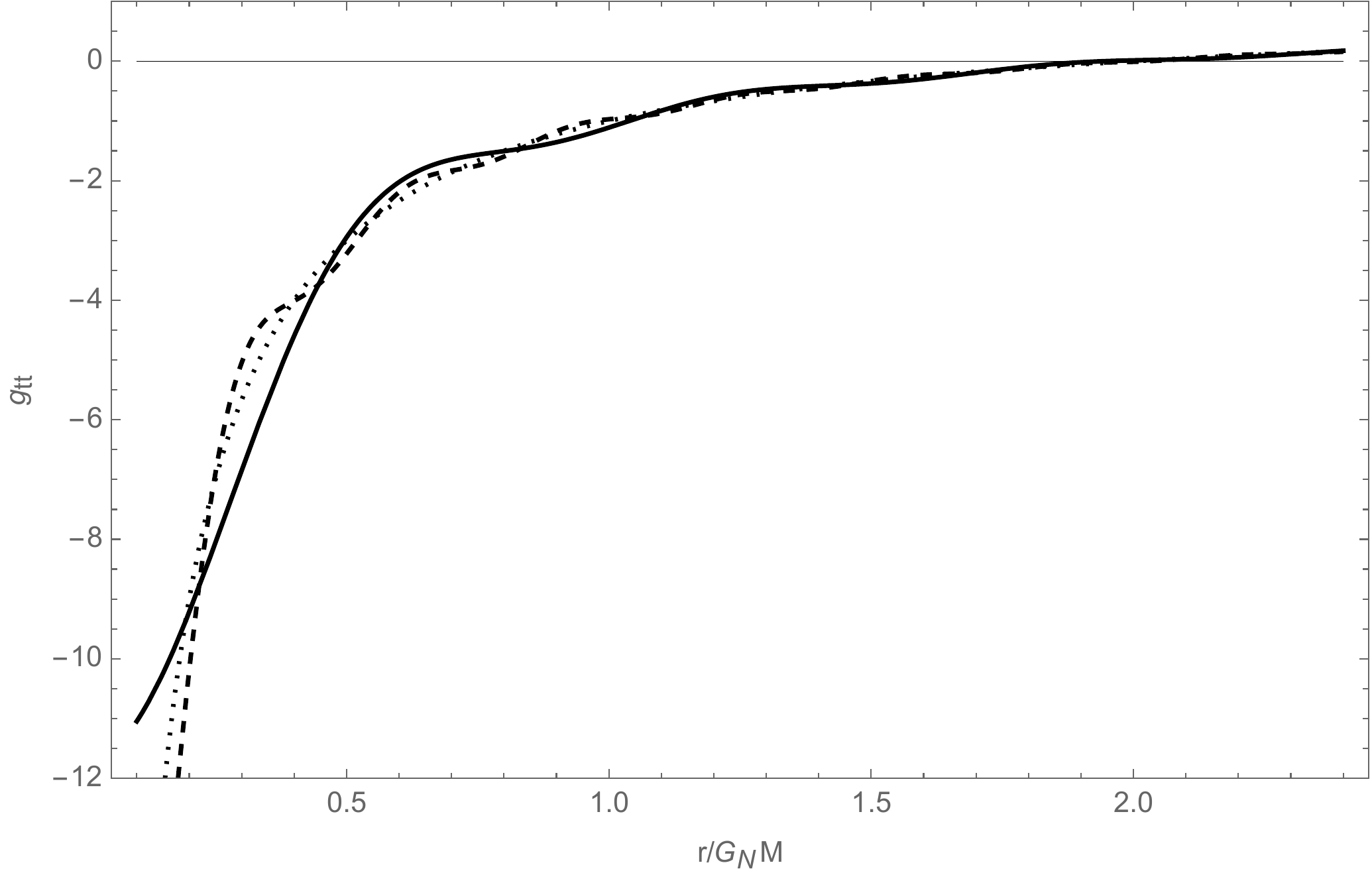}
\caption{Quantum metric component $g_{tt}=1+2\,V_{\rm QN}$ in Eq.~\eqref{gQ} compared to $g_{tt}=1+2\,V_{\rm N}$
(dotted line) for $R=\Rh/10$ (solid line) and $R=\Rh/20$ (dashed line).}
\label{f:gtt}
\end{figure}
The outer spacetime metric can now be written using the effective quantum potential
from Eq.~\eqref{Vq} as~\cite{Casadio:2021eio}
\be
\d s^2
=
-\left(1+2\,V_{\rm QN}\right)\d t^2
+
\frac{\d r^2}{1+2\,V_{\rm QN}}
+
r^2\,\d\Omega^2
\ .
\label{gQ}
\ee
Since $V_{\rm QN}$ is a function of $R$, this result can be directly interpreted as a quantum violation
of the no-hair theorem~\cite{Calmet:2021stu}.
The metric component $g_{tt}=1+2\,V_{\rm QN}$ is plotted in Fig.~\ref{f:gtt}, where one can see that the
quantum corrected metric remains finite for $r\to 0$ and oscillates around the classical expression for
sufficiently large $r$.
These oscillations are in fact a mere consequence of the use of a hard ultraviolet cut-off, but the amplitude
of such oscillations is still indicative of the size of the deviations from the classical geometry.
Near the centre $r=0$, the metric function is approximately given by
\be
V_{\rm QN}
\simeq
-\frac{2\,\gn\,M}{\pi\,R}
\left[
1-\frac{\pi\,r^2}{18\,R^2}
\right]
\ ,
\ee
which is a bounded function with a derivative that vanishes in $r=0$. 
One can conclude that this quantum effective gravitational potential leads to forces that remain finite
everywhere.
\par
The above result is confirmed by the form of the Kretschmann scalar for the quantum corrected metric
in the limit $r\to 0$, which reads
\be
R_{\alpha\beta\mu\nu}\, R^{\alpha\beta\mu\nu}
\simeq
\frac{64\,\gn^2\,M^2}{\pi\,R^2\,r^4}
\ .
\ee
This expression differs significantly from that for the classical Schwarzschild metric~\eqref{gSch},
namely $R_{\alpha\beta\mu\nu}\, R^{\alpha\beta\mu\nu}\sim r^{-6}$ in the same limit $r\to 0$. 
As a result the radial tidal forces are finite all the way into the centre. 
Another way of reaching the same conclusion is by looking at the relative acceleration
between radial geodesics separated by $\delta r$, to wit
\be
\frac{\ddot{\delta  r}}{\delta r}
=
-R^1_{\ 010}
\simeq
\frac{8\,\gn^2\,M^2}{9\,\pi^2\,R^4}
\left(1-\frac{\pi\,R}{4\,\gn\,M}\right)
\ .
\ee
In the Schwarzschild spacetime, where the ``spaghettification'' of matter approaching
the central singularity is supposed to happen, the corresponding relative acceleration
is ${\ddot{\delta  r}}/{\delta r}\sim r^{-4}$ and the difference between the two is obvious. 
\par
In the quantum corrected geometry, $r=0$ is an integrable singularity~\cite{lukash}. 
Some geometric invariants still diverge in this case, but with no harmful effects to matter,
as we have just shown.
Moreover, the Einstein tensor $G_{\mu\nu}$ obtained from the metric~\eqref{gQ} can be used
to calculate the effective energy-momentum tensor $T_{\mu\nu}$ which, in turn, leads to the
effective energy density 
\be
\rho
=
-\frac{G^0_{\ 0}}{8\,\pi\,\gn}
=
\frac{M}{2\,\pi^2\,r^3}\,
\sin\!\left(\frac{r}{R}\right)
\ ,
\ee
as well as the effective radial pressure 
\be
p_r
=
\frac{G^1_{\ 1}}{8\,\pi\,\gn}
=
-\rho
\ee
and the effective tension
\be
p_t\
=
\frac{G^2_{\ 2}}{8\,\pi\,\gn}
=
\frac{M}{4\,\pi^2\,r^3}
\left[
\sin\!\left(\frac{r}{R}\right)
-
\frac{r}{R}\,\cos\!\left(\frac{r}{R}\right)
\right]
\ .
\ee
The integrals of both, the density and the pressure, over all space are finite with the following results
\be
4\,\pi\,\int_0^\infty
r^2\,\rho(r)\,\d r
=
-4\,\pi\,\int_0^\infty
r^2\,p_r(r)\,\d r
=
M
\ee
and
\be
4\,\pi\,\int_0^\infty
r^2\,p_t(r)\,\d r
=
\frac{M}{2}
\ ,
\ee
which explains the integrable quantum nature of the metric.
\par
The quantum corrected metric still contains a horizon when the cut-off scale $R\lesssim \Rh$.
The horizon radius $r=\rh$ is given by the solution of $2\,V_{\rm QN}=-1$, which can be computed numerically
for different values of $R$.
A remarkable result is that the horizon radius $\rh$ shrinks for $R$ approaching $\Rh$.
In fact, $\rh=R$ for $R\simeq 0.6\,\Rh$ and $\rh$ would further vanish for $R\simeq 0.7\,\Rh$.
This means that the material core cannot be too close in size to the classical gravitational radius $\Rh$
in order for it to lie inside the actual horizon.
Moreover, the quantum corrected geometry does not contain an (inner) Cauchy horizon
(when an outer event horizon exists).
Therefore, some potentially serious casual issues that appear for regular black holes~\cite{regular,Carballo-Rubio:2021bpr}
are avoided in this scenario.
Remarkably, this result extends to electrically charged black holes~\cite{Casadio:2022ndh} and can be further
generalised for rotating systems~\cite{Casadio:2023iqt}.
\par
With enough experimental precision, the deviations of the quantum metric from the classical geometry
could be observed on test bodies located outside the horizon at $r>\rh$.
The scale of these deviations depends on the (quantum) size $R$ of the matter source.
Outside the horizon (for $r>\rh\sim \Rh$) their amplitude decreases with the decrease of the ratio $R/\Rh$
(see Fig.~\ref{f:gtt}). 
Therefore, these departures of $V_{\rm QN}$ from  $V_{\rm N}$ become too small to be measured by a distant
observer for small enough (but finite) values of the ratio $R/\Rh$.
(They should also converge to the expected amplitudes obtained from the effective field theory approach
in the weak-field regime.)
This effect can also be interpreted as a damping of transients, which happens as the source radius $R$ shrinks
and the collapse continues behind the horizon until $R\simeq R_M$.
\par
The departure of $\rh$ from the standard Schwarzschild radius $\Rh=2\,\gn\,M$ will also result
in modifications of the horizon area $\mathcal A_{\rm H}$, which ultimately means changes
of the Bekenstein-Hawking entropy~\cite{bekenstein}
\be
S_{\rm QBH}
=
\frac{\mathcal A_{\rm H}}{4\,\lp^2}
=
\frac{\pi\,\rh^2}{\lp^2}
\label{S_BH}
\ee
and black hole temperature~\cite{hawking}
\be
T_{\rm Q}
=
\frac{\hslash\,\kappa}{2\,\pi}
=
\frac{\hslash}{2\,\pi}\left.\frac{\partial V_{\rm QN}}{\partial r}\right|_{r=\rh}
\ ,
\ee
where $\kappa$ is the horizon surface gravity (see Ref.~\cite{Casadio:2021eio} for more details).
Such changes will affect the Hawking emission in a way that depends on the state of the inner
quantum core~\cite{Calmet:2021stu}.
\section{Outlook} 
\label{S:conc}
\setcounter{equation}{0}
In the previous sections we provided a comprehensive scenario supporting the conclusion that black holes
are indeed quantum objects that cannot be described accurately within the effective field theory of gravity.
In particular, the collapsed matter core displays a quantised surface areal radius in qualitative agreement 
with Bekenstein's area law and the same result is obtained by requiring that the outer geometry be
derived as the mean-field approximation of a coherent quantum state.
\par
One of the features stemming from the latter result is that the effective energy density inside the core
behaves like $\rho\sim r^{-2}$ for $r\to 0$, which would be usually ruled out as a classical regular distribution.
However, if one considers that one usually has $\rho\sim |\Psi|^2$ in quantum physics, where $\Psi=\Psi(r)$
is a wavefunction in position space, the only fundamental requirement is the integrability of the quantum state,
that is
\be
\int_0^r
\rho(x)\, x^2\,\d x
\sim
\int_0^r
|\Psi(x)|^2 x^2\,\d x
<\infty
\ ,
\ee
for all $r>0$.
This condition leads to a mass function $m\sim r$ near the centre, which shows that $r=0$ does not
contain any singular source.
\par
A classical energy density is supposed to vanish in $r=0$, which instead implies that $m\sim r^3$.
There is of course no singular contribution at $r=0$ in this case either, but it becomes generically impossible
to avoid the presence of a Cauchy horizon when the event horizon exists.
Remarkably, the quantum nature of black holes with $m\sim r$ could solve this issue as well as removing the
central singularity, at least in spherically symmetric configurations~\cite{Casadio:2021eio,Casadio:2022ndh}.
\par
Not surprisingly, for axially symmetric (rotating) black holes, the case is more involved.
Assuming a constant specific angular momentum $a=J/M$ inevitably brings back the inner horizon,
even if the mass function $m\sim r$ and there is no central singularity.
However, there is no reason to not allow for $a=a(r)$ in such a way that it vanishes sufficiently 
fast for $r\to 0$ to again avoid the inner horizon~\cite{Casadio:2023iqt}.
\par
From all of the above preliminary considerations, it looks promising to develop a more refined
description of black holes as quantum objects which are safe from pathologies like the central
singularity and Cauchy horizons, and still reproduce the well-tested geometries in the exterior.
Of course, to validate this picture, one must find testable predictions that deviate from the
classical general relativity phenomenology.
The outer hair of these black holes is the natural candidate for further investigations.
%
%
%
%
\begin{acknowledgement}
R.C.~is partially supported by the INFN grant FLAG and his work has also been carried out in
the framework of activities of the National Group of Mathematical Physics (GNFM, INdAM).
O.M.~was supported by Romanian Ministry of Research, Innovation and Digitalisation under
Romanian National Core Program LAPLAS VII - contract no.~30N/2023.
\end{acknowledgement}
%
%
%


\end{document}